# SELF-REGULATING SURFACES FOR EFFICIENT LIQUID COLLECTION

Christian Machado[1†], Yuehan Yao[2†], Emma Feldman[3], Joanna Aizenberg[4,5*], and Kyoo-Chul Park[1*]

[1]*Department of Mechanical Engineering, Northwestern University, Evanston, IL, 60208*
[2]*Department of Materials Science and Engineering, Northwestern University, Evanston, IL 60208*
[3]*Department of Chemical Engineering, Northwestern University, Evanston, IL 60208*
[4]*Wyss Institute for Biologically Inspired Engineering, Harvard University, Cambridge, MA 02138*
[5]*John A. Paulson School of Engineering and Applied Sciences, Harvard University, Cambridge, MA 02138*
[†]*Contributed equally*

ABSTRACT:

To achieve efficient liquid collection, a surface must regulate incoming liquid accumulation with outgoing liquid transport. Often, this can be proposed to be achieved by functionalizing surfaces with non-wetting characteristics. Yet, there remain fundamental, practical limits to which non-wetting surfaces can effectively be employed. We instead utilize filmwise wetting to achieve liquid regulation via a Laplace pressure gradient induced by solid surface curvature. The key parameters affecting this capillary flow are then introduced, namely solid properties like scale and curvature and liquid properties like surface tension and density. The liquid regulation mechanism can then be employed in condensation and aerosol processes to generate enhanced flow, while the solid geometry needed to create this capillary flow itself is capable of affecting and enhancing liquid generation. Ultimately, the surface design framework can be customized to each unique application to optimize processes in HVAC, industrial steam generation, chemical depositions, and atmospheric water harvesting.

MAIN TEXT:

Efficient liquid collection is essentially a problem of surface regulation. Incoming liquid onto the surface must balance liquid transporting off the surface to achieve no net accumulation. The generation of incoming liquid can occur in multiple ways: (*i*) by nucleating a condensate from a vaporous source, or (*ii*) impacting an already condensed particle or droplet from an aerosol flow. Then, why is it advantageous to achieve no net accumulation? Because liquids can be thermally insulating (much more than the typically metallic substrates they deposit on)[1], can cause instabilities with passing airflows, resulting in increased drag[2-4], and can re-integrate into a previously separated medium[5-7]. If the species is water, these issues can be embodied in HVAC, in aviation, and in atmospheric water harvesting. While one workaround is just to remove the liquid generation term, there are many applications that that possibility is either impractical (*i.e.*, air conditioning and dehumidification) or undesirable (*i.e.*, atmospheric water harvesters). Therefore, the major strategy to achieve self-regulation is to maximize the outflow of liquid from the surface, once it is deposited.

Researchers have addressed this scientific question by engineering surface patterns across various length scales to modify a substrate's wetting[8-13]. Generating surfaces with greater non-wetting characteristics reduces interfacial adhesion and increases the mobility of liquid droplets on the surface. Yet, this mobility can be largely uncontrolled, both with in-plane motion and out-of-plane droplet ejection[14]. With enhanced transport, though, condensation on these engineered surfaces can be enhanced, but in a practically limited way. They apply mainly to water-based systems, as generating non-wetting behavior is facilitated by increased surface tension. With lower surface-tension liquids, non-wetting, Cassie drops are much more challenging to attain[15]. So is also the case with condensation in large degrees of supersaturation[16,17]. In the event of failure, liquid impingement of the engineered textures not only negates the enhanced transport, but, in fact, makes liquid transport more tortuous. Aerosol deposition on these engineered surfaces is similarly plagued by low surface-tension droplet depositions, which negates most traditional techniques to promote Cassie droplets. But, they are also further plagued by re-entrainment, because air drag can overcome the solid-liquid interfacial adhesion (which, for non-wetting surfaces, is quite low) and displace particles back into the airflow[5,7]. So then, if air drag should be constrained to being less than the interfacial adhesion, the result should be engineered surfaces with less mobile droplets (in order to increase adhesion such that $F_{adh} > F_{drag}$). This, paradoxically, reduces the surface's capability for self-regulation, though.

Overlooked is the potential to create a self-regulating mechanism by *increasing* surface wetting. Filmwise, or highly wetting, surfaces are generally undesirable for enhancing liquid transport because of their high solid-liquid adhesion, associated with their large solid-liquid interfacial area. This typically results in: (*i*) liquid films promoting a high degree of thermal resistance as compared to discrete droplets in condensing systems[1,18] and (*ii*) liquid films contributing especially to the clogging of permeable surfaces in aerosol deposition systems[5,19]. These are significant issues, but are problematic because the thickness of the liquid films goes unregulated by external forces. However, the failure mechanisms for dropwise surfaces do not exist for filmwise, namely surface tension, particle size, and airflow velocity, and so the potential for their expanded use is far greater, if only they could self-regulate their films such that they are and remain negligibly thin.

To create a passive, self-regulating mechanism, we utilize both capillary and gravitational shedding to reduce the liquid residence time on the surface and promote very thin films. Initial inspiration was derived from the *Welwitschia mirabilis*, a unique plant in the Namib Desert with a lifespan of several thousands of years[20]. While the entire plant's function is incredibly complex, we instead choose to focus on the structure of its leaves, which have a striking oscillatory pattern not commonly observed with other biological organisms. Using scanning electron microscopy (SEM), the leaves reveal a multi-scale texture that exist both on the millimeter and nanometer length scales. From this, a simplified, periodic wavy structure is designed such that the peaks have a positive curvature, the valleys have a negative curvature, and a linear curvature gradient exists between the two locations, as shown in Fig. 1a. The surface design is easily transferrable to a thin, aluminum sheet via compression molding, then nanotextured via Boehmitization to promote multi-scale heirarchies[21].

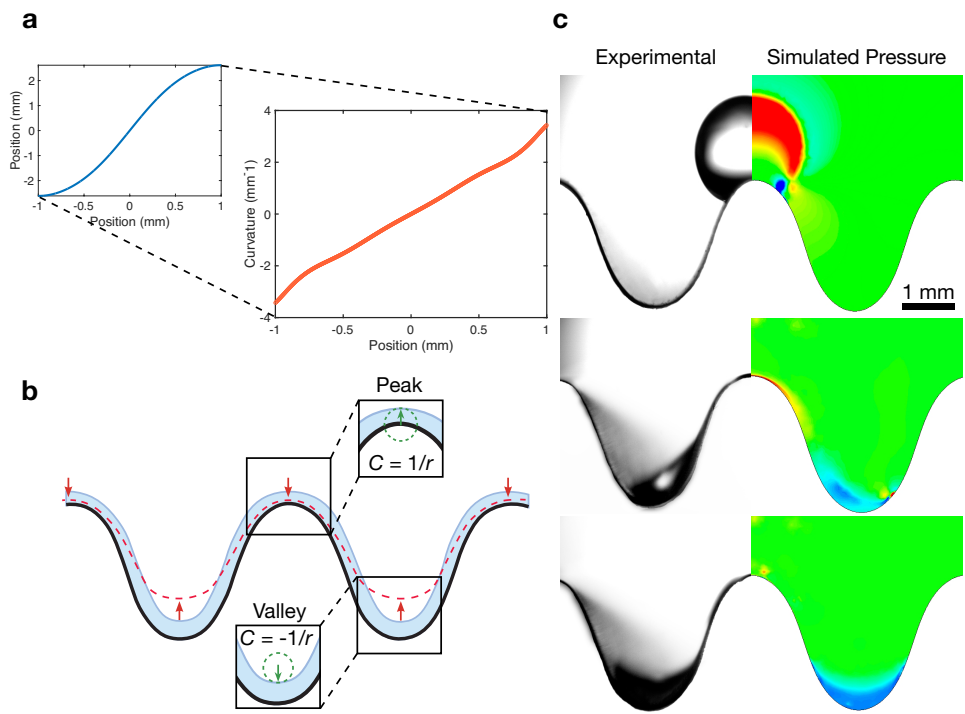

**Figure 1. (a)** A plot of the surface profile used to generate the wavy surfaces for this study and its corresponding curvature. **(b)** A schematic of the liquid dynamics on this surface. A consequence of the substrate's curvature gradient is that it induces a pressure gradient within the liquid when it takes on the curvature of the substrate beneath it. **(c)** When the liquid takes on this curvature gradient during spreading, an internal pressure gradient forms within the film that drives liquid transport from peak to valley.

By effectively creating this hierarchical surface roughness, liquid interacting on the surface behaves in a Wenzel state, associated with a very low contact angle, and high spreading potential[8]. If, then, a water droplet is deposited on the peak of a wave, the liquid is immediately wicked into the valley, as visualized by the schematic in Fig. 1b. Conversely, if the wavy surface is designed to promote a dropwise state (by coating the aluminum surface with a hydrophobic polymer), then no such motion is observed and a droplet deposited on the wave's peak remains there, unperturbed. This fluid flow is simulated using a finite element method to visualize an asymmetric pressure distribution within the liquid film, as shown in Fig. 1c alongside an experimental comparison. This pressure distribution is primarily a result of a Laplace pressure gradient that forms when the liquid film takes on the highly variable curvature of the solid surface beneath it. As it is, the wavy peak corresponds with a positive curvature and the valley with a negative curvature, resulting in the liquid film at the peaks having a higher pressure than at the valleys. If the liquid in unable to spread (*i.e.*, when drops form rather than films), then a negligible Laplace pressure gradient is easily negated by contact line pinning.

Pressure-driven flow will cause the liquid film at the peak to drain, causing the film thickness there to rapidly decrease. As the film thickness continues to decrease to below ~ 100 nm, the disjoining pressure becomes relevant, largely in part, due to the increasing magnitude of short-range van der Waals forces. The hydrophilic interactions between liquid, solid, and vapor produce repulsive pressures that offset the initial Laplace pressure gradient flow at very thin films[22]. Assuming that van der Waals repulsion is the dominant component of the disjoining pressure, then equating that with the capillary pressure of the liquid film (the hydrostatic pressure can be neglected here because the film thickness is so small) generates a prediction for the thickness of the remaining liquid film at the wavy peak[23], as such:

$$B = -\left(\frac{A_H R}{6\pi\gamma_l}\right)^{1/3} \quad [1]$$

Here, $B$ is the film thickness, $A_H$ is the Hamaker constant, $R$ is the radius of curvature of the liquid (in this orientation, only one radius of curvature is accounted for as the second radius approaches infinity), and $\gamma_l$ is the liquid surface tension. The Hamaker constant for aluminum-water-vapor is approximately on the order of $10^{-20}$ and the radius of curvature of the film is assumed to be the radius of curvature of the solid substrate. Therefore, the stable film thickness is plotted against curvature, as shown in Fig. 2a (again, assuming that the van der Waals interactions are the dominant surface forces). We consider surface forces to be non-negligible below film thicknesses of ~100 nm[24], meaning that Eq. 1 should be valid for radii of curvature of ~ 4 cm, or around one order of magnitude greater than the wavy surfaces examined here. For the specific peak radii of curvature here, a calculated, idealized stable film thickness of around 20 nm is expected (although in reality, there are additional electrostatic and oscillatory surface forces that can contribute to metastability).

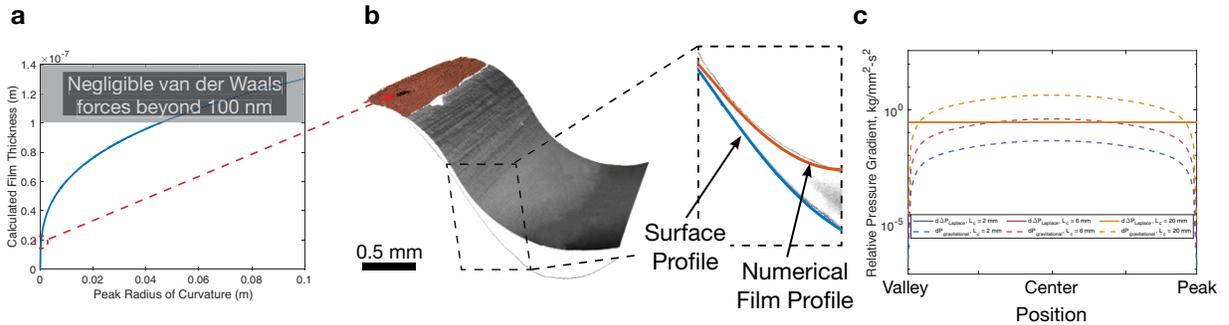

**Figure 2. (a)** Assuming the van der Waals forces are the dominant component of the Disjoining pressure, then equating the Laplace and Disjoining pressures should reveal that the static film thickness at the peak should scale as $r^{1/3}$. **(b)** A 3-dimensional construction of the liquid profile using a laser confocal. The inset shows a cross-section and overlaid with a numerical prediction of the film profile. **(c)** A plot comparing the relative dominance of either capillary or gravitational forces to induce liquid motion at different scales.

The liquid film was then experimentally visualized using laser confocal microscopy (Leica SP8, 10x objective in air). To reduce evaporation during imaging, a solution of 60% glycerol and 40% water was made, similar to a previous study[25] and to promote fluorescence at 405 nm excitation, fluorescein dye was added. As shown experimentally, a thin liquid film extends nearly to the wave's peak, until fluorescent activity is no longer detected (indicating either film rupture or film thickness below the detection limit). While extremely challenging to visualize

The equilibrium film profile should exist when the capillary pressure is balanced with hydrostatic and disjoining pressures, written as:

$$P = \gamma C(x) - \rho g z(x) - \frac{A_H}{6\pi z(x)^3} \qquad [2]$$

where $C(x)$ is the curvature gradient of the liquid film-air interface, and $z(x)$ is the thickness of the liquid film anywhere along $x$ in the domain. The interfacial curvature function can be expressed as a differential equation of the film thickness, $z(x)$, as such:

$$C(x) = \frac{z''(x)}{(1+z'(x)^2)^{3/2}} \qquad [3]$$

This equation represents purely the curvature of the liquid film, but in this case, the total curvature of the liquid film-air interface is the sum of the film's curvature and the curvature of the solid substrate beneath. In a similar way, the hydrostatic pressure is influenced by the substrate curvature, and so the film thickness is normalized to the unit component that is parallel with gravity. The complete differential equation is solved numerically, and overlayed on the experimentally obtained film profile in Fig. 2b.

At the length scale of the wavy feature tested, capillary flow dominates the surface dynamics, as evidenced by peak-to-valley film transport regardless of gravitational orientation. As the length scale gets larger, or the curvature gradient gets smaller, though, it could be predicted that the driving force for liquid transport could transition from capillary to gravitational. This is important to understand because, for gravitationally driven flow, a required orientation to achieve self-regulatory flow is imposed. The driving force for capillary transport is the gradient of the Laplace pressure of the liquid film[26]. Assuming the liquid film at least instantaneously takes on the profile of the solid substrate, then the Laplace pressure gradient, in this case, is linear (due to the design of the solid surface profile). Adjusting the period and amplitude of the wavy profile then affects the magnitude of the gradient. The capillary driving force can be expressed as:

$$\nabla(\Delta P_{Laplace}) = \frac{\gamma_l}{r^2}\left(\frac{dr}{dx}\right), \quad where \quad \frac{1}{r(x)} = \frac{y''(x)}{(1+(y'(x))^2)^{3/2}} \qquad [4]$$

Gravitational forces will be most significant when directed opposite the direction of capillary-based fluid flow (*i.e.*, in an upside-down orientation). These gravitational forces depend on the vertical component of the liquid film, which can be approximated by the vertical component of the solid profile, again assuming the liquid film at least instantaneously takes on the solid profile. This component can be expressed as:

$$\nabla(P_{gravitational}) = \rho_l\, g\, y'(x)\, sin[arctan(y'(x))] \qquad [5]$$

In Equations 4 and 5, $\gamma_l$ is the liquid surface tension, $\rho_l$ is the liquid density, $g$ is the gravitational acceleration, and $y(x)$ is the profile of the solid surface. Assessing the relative magnitudes of each expression helps to predict which mechanism is mainly responsible for fluid flow. From this, it is clear that $y(x)$ has an impact on this inequality. Increasing the gravitational component can be accomplished in two ways: (*i*) $y(x)$ can be tuned such that the wavelength decreases (while conserving the amplitude), making the vertical component of the surface larger or (*ii*) uniformly scaling up the feature (both amplitude and wavelength). With the former, increasing the ratio of amplitude to wavelength not only increases the gravitational component, but also the capillary component due to an increase in the curvature gradient. Instead, uniformly increasing the scale of the features has a much greater impact, as shown in Fig. 2c. As the length scale of the feature approaches and exceeds the capillary length of the liquid (in this case, water), gravity plays an outsized role in the film dynamics, stipulating a critical design threshold. Surface tension and density will also play roles, but so long as the feature length remains below the capillary length, capillary flow will still dominate (although the ratio of capillary to gravitational forces might change).

Because of the capillary dominance for small-scale features, the film dynamics are largely independent of orientation, making self-regulation for liquid collection processes far more universal. The implications of this structure on condensation are three-fold. Firstly, capillary flow creates a controllable transport mechanism for condensing droplets when the surface is highly wetting, as shown in Fig. 3a. This helps to promote self-regulation by enhancing transport away from regions of potential liquid generation. Secondly, condensing liquid will result from a concentration gradient near the surface that is made non-uniform by the same millimetric wavy structures that promote capillary flow[27-29]. A non-uniform flux field is the byproduct of a variable concentration gradient across the domain, resulting in locally focused vapor flux at the peaks and nearly completely suppressed vapor flux in the valleys, as shown in Fig. 3b(i). This suggests that the condensation rate will be highest at the peaks, the same region that was designed to promote liquid transport away. With spontaneous liquid transport here, the outflow of liquid can resemble the generation of incoming condensate, reducing accumulation and, in turn, reducing the thermal insulation that scales with average liquid film thickness. Further yet, the liquid transport mechanism directs most of the liquid into the valley region. During condensation, this is of little consequence because the diffusion flux is minimized there, so the effects of increased thermal resistance are minimal. More importantly, these surfaces serve to confine liquid to where it is less susceptible to evaporation, as shown by the simulation of evaporative flux in Fig. 3b(ii).

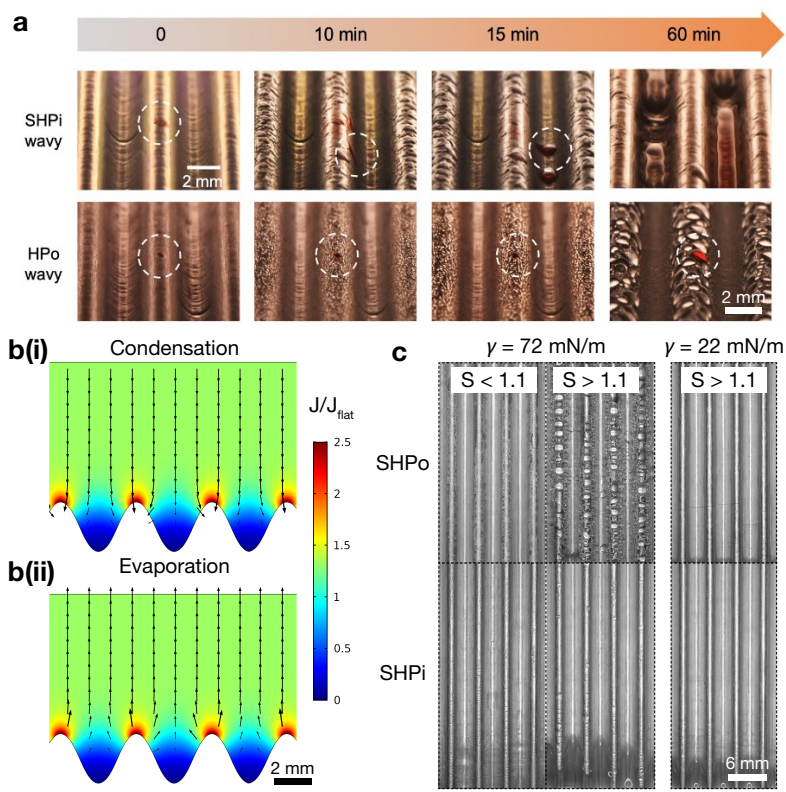

**Figure 3. (a)** Experimental validation of the liquid regulation mechanism being utilized in a condensation environment. A small amount of dye was added to the wavy peak, and the dyed liquid was then tracked over time for superhydrophilic (SHPi) and hydrophobic (HPo) surfaces. **(b)(i)** A finite element simulation of diffusion flux, showing the peaks exposed to focused diffusion flux, suggesting that is where the majority of condensation would occur. **(b)(ii)** A simulation of evaporative flux, showing that the valleys, the regions where liquid is expected to migrate to, are subject to very low evaporative flux. **(c)** Because the condensation dynamics occur in a filmwise wetting regime, the typical points of failure for superhydrophobic surfaces (SHPo) (namely, supersaturation and surface tension) have negligible effect on SHPi surfaces.

While it is the geometry that creates these diffusion patterns, it is the filmwise wetting that promotes self-regulation. Non-wetting wavy surfaces do not trigger capillary transport from peak to valley, and so condensate accumulates where it is mostly generated (the peaks). Therefore, it is actually more desirable to promote filmwise wetting here, which ultimately has broader implications. Cassie-Baxter wetting generally fails in two main ways: (*i*) increasing supersaturation and (*ii*) decreasing the surface tension of the condensing species, both causing liquid impingement of the nanotextures and increasing solid-liquid adhesion. Then, if the purpose of Cassie-Baxter surfaces is to achieve self-regulation by reducing adhesion, then liquid impingement directly affects the ability to regulate liquid accumulation. Instead, these filmwise wavy surfaces promote self-regulation not by reducing adhesion, but by increasing the driving force, a distinction that is largely resistant to changes in supersaturation and surface tension. This is shown experimentally in Fig. 3c by visualizing the wetting transition that occurs for traditional superhydrophobic surfaces, in stark contrast to the uniform wetting characteristics of filmwise wavy surfaces.

In a similar vein, it is the combination of geometry and wetting on these wavy structures that demonstrates enhanced deposition and liquid regulation for efficient aerosol/mist collection. Instead of diffusion, though, these impacts are realized via improved deposition of multi-phase flows and enhanced liquid drainage away from the surface. To improve collection efficiency, a surface can/should increase the number of particles directed towards it[6,30], increase the number of particles close by that deposit on the surface[31,32], and quickly remove previously deposited particles to reduce potential losses like evaporation and re-entrainment[33,34].

The wavy surfaces impact both how the air flows around the structure and how deposited liquid flows on the structure. As airflow encounters the wavy structures, flow separation can occur as a function of the Reynolds number and wave amplitude[35]. Depending on the extent of separation, vortex formation within the wavy channels can be predicted and observed, as shown in Figs. 4a and 4b(i). Vortices help to promote droplet ejection out of the fluid streamlines due to the presence of centrifugal forces that can dominate air drag[36]. As such, the magnitude of the vorticity then affects the potential for droplet deposition on the surface, which, in turn, is based on the bulk Reynolds number[37]. Centrifugal droplet ejection therefore represents one main effect of wave-induced flow separation that promotes increased particle depositions. A second effect of this flow separation is re-attachment slightly offset from the center of the peak, which poses an additional, significant source of particle deposition there, as shown in Fig. 4b(ii). Therefore, the combined effects of these two mechanisms improve overall deposition compared with a non-wavy geometry. The implications for this are realized in the overall fog collection rates for these wavy surfaces, as they can demonstrate a roughly *5x* enhancement in collected water over flat surfaces that do not observe flow separation, as shown in Fig. 4c.

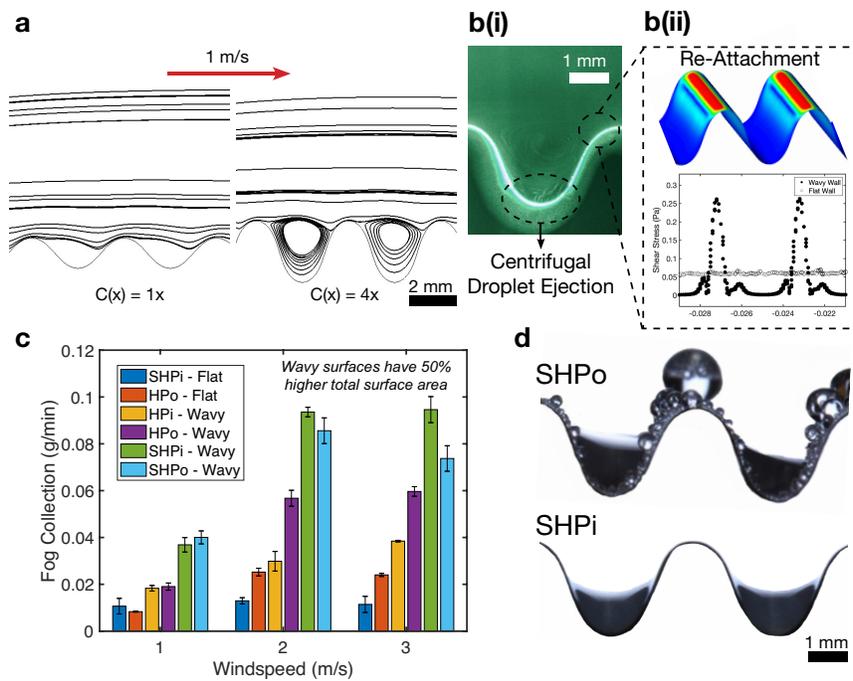

**Figure 4. (a)** Finite element simulations of fluid streamlines visualizing the flow patterns of incoming 1 m/s flow over wavy surfaces of varying curvature gradients. **(b)(i)** Experimental confirmation of the vortex patterns and corresponding visualization of droplet ejection. **(b)(ii)** A simulation of wall shear stress, meant to model the effects of flow re-attachment near the peaks. **(c)** Fog collection performance comparing the effects of wettability and geometry. **(d)** Experimental visualization highlighting the effects of wettability and, therefore, self-regulation, during airflows.

Obvious from these results, additionally, is the impact of surface wetting on fog collection rate, via its effects on liquid drainage. Drainage can be degraded when the onset time of liquid droplets on the collection surface is high, due to an increasing likelihood of evaporation and re-entrainment[33]. Because the filmwise wavy surfaces promote both spontaneous capillary transport into the valleys and gravitational transport downwards, negligible liquid remains on the surface, whereas comparable dropwise wetting surfaces have much longer onset times due to contact line pinning. This becomes very visible in Fig. 4c, where the surfaces that promote more mobile liquid produce the largest fog collection rates (when omitting liquid still on the collector upon experimental completion). At low airspeeds, these differences are largely a result of droplet mobility after deposition rather than differences in deposition based on wetting.

At higher airspeeds, though, there becomes a divergence between the most mobile surfaces: superhydrophobic and superhydrophilic. All surface wettabilities will observe droplet growth where deposition is high: at the flow re-attachment point. Yet, for the non-wetting surfaces, there exists a critical size requirement for gravity to overcome adhesion and shed the droplets, suggesting this cannot occur immediately. In contrast, the superhydrophilic surfaces will undergo spontaneous liquid transport, although primary particle deposition near the peaks is also expected. This effect is shown in Fig. 4d and underscores the influence of surface wettability. Because re-entrainment, or the re-introduction of droplets into the multi-phase flow, is based on the local drag force of the airflow (which scales with velocity) and droplet adhesion (which scales with contact angle hysteresis (CAH))[5], an increased exposure to the bulk airflow and reduction in contact angle suggests much greater potential for droplet re-entrainment. On the contrary, the superhydrophilic wavy surfaces spontaneously wick deposited droplets (regardless of their position on the oscillation) into the channels, where they are effectively shielded from the incoming bulk airflow. Therefore, the most wetting surfaces promote the most efficient fog collection due to enhanced droplet deposition (influenced by the wavy geometry) and a reduction in drainage losses (influenced by the self-regulatory film dynamics).

By utilizing macroscopic curvature gradients, liquid films can become highly mobile and controllable. This can be effectively implemented by creating surface regions of preferred deposition and those of preferred accumulation.  Connecting these regions with capillary flow creates a mechanism for surfaces to self-regulate. We introduce variables than can impact fluid flow and geometries where this fluid flow can be used to great effect. Because geometry can play a role in liquid collection (via diffusion or particle deposition), the surfaces can be tuned to specific applications like atmospheric water harvesting, industrial HVAC, chemical mist elimination, steam power generation, and chemical depositions.